\documentstyle[11pt,DL95]{article}

\begin{document}   

\title{Digital Library Technology for Locating and Accessing Scientific Data}

\author{ Robert E. McGrath, Joe Futrelle, Ray Plante, \\
National Center for Supercomputing Applications \\
University of Illinois, Urbana-Champaign \\
Damien Guillaume, \\
Universit\'{e} Louis-Pasteur, Strasbourg, France}

\maketitle

\begin{abstract} 

In this paper we describe our efforts to bring scientific data 
into the digital library. This has required extension of the
standard WWW, and also the extension of metadata standards far beyond the
Dublin Core. Our system demonstrates this technology for real scientific
data from astronomy.

\end{abstract}


\section{INTRODUCTION}
In the last few years we have seen the evolution of the Internet and WWW
from a loose connection of information sources toward information rich
digital libraries. (e.g., \cite{CICVEL,DLI,Leiner98,LagFi98}) 
In our view, a digital
library is an organized way to locate, access, and analyze digital artifacts
of many kinds, federating many data services to create a virtual information
space.

The evolving digital library may play a key role for scientists by providing
a unified environment for information discovery and access. In particular,
the digital library can go beyond the traditional library, and provide
direct, immediate location and access to both literature and data.

At NCSA, the Emerge project has been constructing the basic infrastructure
required for interoperable searching and for analysis of many kinds of
data from many heterogeneous data services
distributed about the network. \cite{emerge} 
In our current work, we are implementing a prototype
to demonstrate the effectiveness of this technology for searching and accessing
astronomy data. This prototype illustrates how our flexible architectures
can be applied to a collection of existing systems, to create an enhanced
environment for information discovery.
\cite{P30} 
This work has been a collaboration
of astronomers and computer scientists and NCSA, NASA, University of Ulster,
and elsewhere. 
The NCSA Astronomy Digital Image Library (ADIL)
has been the key testbed for demonstrating the technology.
\cite{Plante97,PCM98}

In this paper, we describe our model and prototype implementation
for interoperable search and analysis as applied to scientific
digital libraries.  Our model places constraints on standards
necessary for meaningful interoperability.  In particular, we will
show that the Dublin Core alone is insufficient to support the
metadata associated with complex scientific data.  However, 
appropriate standards can facilitate richer forms of research tuned
to a distributed scientific environment.

\subsection{Digital Libraries for Science}

For scientists, the digital library is particularly important because
of the importance of digital data and analysis, and the need for timely
and rich information exchange. Today, scientific discovery and communication
routinely creates and uses many kinds of significant digital components:
\begin{itemize}
\leftmargin=0.3in
\item
raw data
\item
analyzed data
\item
imagery
\item
analysis environments
\item
simulations
\item
notes, letters, and reports
\item
published articles
\end{itemize}

These digital artifacts are complex and inter-related; for example, a digitally
published article ``points" to the data, instrumentation, and software which
are described and interpreted by the text. Similarly, the digital
representations of simulations and analyzed data such as images are most
useful and valid in the context of the published documentation and scientific
reports. An archive of scientific data will contain pointers from the data
to published articles which explain and validate them; and scientific articles
contain pointers to the data which they report. Similarly, theoretical
results in the form of computational models are intended to be correlated
with relevant observational and experimental data.

There are many repositories of scientific data already on line, and
each new scientific project almost inevitably produces significantly 
larg\-er amounts of digital data. \cite{ISDA}  
Through the use of the World Wide
Web and URLs, scientific information is already becoming a rich web of
connected digital information. However, it remains a significant
and lasting challenge for humans to exploit this richness, to discover,
access, and understand the knowledge that may reside or be created from
digital resources. We believe that these archives should be an integral
part of the digital library of the future, bringing together all types
of scientific information resources in a single environment.

The Emerge project at NCSA is developing practical infrastructure for this
new type of digital library.
In this vision, a student or researcher could ``go to the library" to
ask a scientific research question. For example, a researcher could
seek to inquire about the climate in Illinois in recent years. Even
today, the library would provide pointers to published literature about
weather, vegeta\-tion, wild\-life, and so on, much of which is
available on-line. 
The results should also provide
pointers to relevant climate data, satellite imagery, computational 
models, and resources such as email archives. 
In most cases, these resources already
exist and are available on the Web, but locating and accessing this diverse
set of materials would be difficult without the organizing and facilitating
role of a new kind of research library. This kind of digital library
will not only make routine scientific information finding more efficient,
it will enable cross-discipline and synergistic discovery; 
since the investigator
will likely be presented with information from many unexpected sources.

\subsection{A Case Study: Astronomy and Space Science Data}

In recent decades, we have experienced a golden age for the exploration
of the universe. 
New ground and space based instruments and
powerful computing systems have produced an explosion of astronomy and
space science data. 
This explosion has driven the development of
data archives, digital libraries, and other network-based services that
make it easier to access research-quality information. 
The
success of such services has created
environments within which one can gather knowledge
from diverse sources to address new scientific questions.

\subsection{The Astronomy Digital Image Library (ADIL) Testbed}

The NCSA Astronomy Digital Image Library \newline (ADIL) has been the key testbed
for demonstrating the technology. The ADIL was developed with support
from NASA and the National Science Foundation to address some of the challenges
of distributing scientific data over the network. 
\cite{Plante97,PCM98}
Its specific
mission is to collect fully processed astronomical images in FITS format
(a standard astronomical image format \cite{FITS}) and make them available to
the research community and the interested public via the World Wide Web.

The ADIL allows users to search, browse, and download
astronomical images. As we will discuss below, this 
can be a non-trivial process when the images are not in
the usual GIF or JPEG formats.

The ADIL is more than a tool for astronomers looking for images to augment
their research. It is also a means for authors who wish to share
their images with the community. While many of the Library's images
come from observatories, the core of the collection comes from individual
authors. The ADIL provides a way to upload the images to the Library,
along with any supporting data, where it can be processed and made available
to the Library users.

Authors deposit images into the Library in the form of collections we
refer to as ``projects". Normally, an author would make a deposit
at the end of some scientific study when the resulting publication is going
to press; all the fully processed images associated with that paper would
make up the project. In this way, the ADIL is part of the new paradigm
for scientific publishing. 
\cite{Plante97,PCM98}

\section{EXTENDING THE WWW MODEL: A `CONVERSATION WITH THE DATA'}

In the conventional WWW model, which is biased toward small, text-oriented
documents, a data location service usually returns a set of URLs pointing
to documents which the user must visit--i.e. download to the client--to
view and analyze. For scientific data, ``search-and-download" is not
a practical model because the objects are typically not "documents", but
rather large, complex objects (datasets) stored in formats not supported
by standard browsers (such as FITS \cite{FITS} or HDF \cite{HDFHome}). 
In earlier work, we 
described the need for a ``conversation with the data'' which 
extends the standard
Web model. 
\cite{FM96,REM97}
The basic scenario is:

\begin{enumerate}
\itemindent=0.3in
\listparindent=0.3in
\item
search to locate candidate data objects
\item
browse and select the objects
\item
download selected data for further analysis
\end{enumerate}

Scientific archives have adapted to the Web by integrating a browsing stage
to the information discovery process (e.g. \cite{Plante97,Yeager97}). 
In so-called {\em server-side
browsing}, the data provider presents a preview of a dataset which might
include a GIF or JPEG rendering of the data and a display of some subset
of the associated metadata, all packaged as an HTML document.  The ADIL
is a good example of such a data service.\cite{ADILExample}

This model can work very well when interacting with a single data provider
and a set of datasets that isn't too large. However this model becomes
quite laborious to the user when trying to interact with more 
than a few weterogeneouwith s
data providers because:
\begin{enumerate}
\itemindent=0.3in
\item
a single question must be entered differently into each of the providers'
custom interfaces
\item
each HTML-formatted query response and associated browsing documents
must be visited for visual interpretation in a series of individual, stateless
requests, e.g., a list of links to URLs.
\item
browsing of the data items is limited to what is provided 
by each data providers' interfaces.
\end{enumerate}
Also, some kinds of user interaction are difficult to implement with server-side
browsing. For instance, drawing a bounding box or dynamically fiddling
with color maps is difficult to implement well on a server.

To address these issues we extend the ``search-browse-and-download" model
by adding:
\begin{enumerate}
\itemindent=0.3in
\item
stateful communication with a data-provider,
\item
support for standard query profiles,
\item
support for standard record format appropriate for scientific data.
\end{enumerate}
The result is a more fluid, automated, and efficient interaction with multiple
data providers. Users can interact with data from one provider while
queries to other providers are being processed. The browsing can
occur with different levels of detail. Perhaps the most powerful
feature is that clients can take greater control of the browsing by plugging
in specialized visualizers for quick plotting or manipulating of the results.

\section{INFRASTRUCTURE FOR INFORMATION DISCOVERY FOR SCIENTIFIC RESEARCH}

Information discovery is increasingly the most critical component of scientific
research. 
As scientists work to solve problems, they need multi-modal
access to geo\-graphically-distributed collections of large and highly structured
data sets. 
Discovering which data sets are potentially relevant to
a particular problem involves more or less elaborate characterizations
of the data in terms of domain-specific attributes. Furthermore,
examining candidate data sets to locate the most relevant ones involves
highly specialized interactions with the data.

Yet this diversity must not come at the cost of interoperability.
For science to progress, it is crucial that scientists be able to locate
information from many different scientific domains when attempting to solve
a problem in their own domain. Cross-disciplinary researchers should
not be burdened with a different set of information discovery software
tools for each discipline they work in, especially in the respects that
those software tools perform essentially the same functions. Also,
data services should provide data not only to individual end users but
also to other services which add value to them.

Even within a single discipline, it may be necessary to query many data
repositories in order to locate all the data relevant to a scientific question.
For instance, the NASA Space Science Data System Technical Working Group
reports a real scenario based on the investigation of sulfur ($S_2$) on comets.
This investigation turned out to require data from multiple sources, including
several spacecraft, ground based telescopes, the Hubble Space telescope,
and published (and unpublished) scientific literature. 
The data was retrieved from
many different sources, in widely different formats.
(\cite{SSDS}, Appendix 2)

A distributed information discovery infrastructure should be built which
emphasizes standard search protocols, file formats and general purpose-tools.
It should be designed in such a way that profiles, formats, and browsers
specific to a particular domain can be easily plugged in to the infrastructure
and shared between data providers and consumers.

In a sense, the WWW already provides a semblance of such an infrastructure.
HTTP supports a variety of file formats and forms-based CGI services can
be used to implement search tools which return views of information with
additional forms controls for manipulating the view. However this
mode of using the WWW does little to advance a standard query syntax or means
of defining metadata schemas and profiles. Also, it fails to separate
the user interface for information retrieval (the HTML forms) from the
delivery of information itself (the metadata in the page). Search
results returned as part of an HTML page are not standardized and cannot
be easily be compared to similar results from a different service.

\subsection{NCSA Emerge: Practical Infrastructure For Information Discovery}

The NCSA Emerge Project is addressing these issues, with the goal of
designing infrastructure to create unified information discovery across
heterogeneous databases; 
developing free software based on standards (Z39.50 \cite{Z39.50},
XML \cite{XML}). The Emerge software includes \cite{emerge}:
\begin{itemize}
\itemindent=0.3in
\item
the {\em Gazelle} gateway, which adds Z39.50 to a database (using Z39.50 is
recommended but not required)
\item
the {\em Gazebo} search gateway, which manages searches across multiple heterogeneous
databases
\item
a Java client toolkit, which communicates with the Gazebo gateway, creating
queries and presenting results.
\end{itemize}
This infrastructure is being developed for several applications, including
engineering literature \cite{DLI} and medical research databases, 
as well as astronomy.

\subsection{Profiles}

In order to build a distributed search infrastructure which serves the
scientific community, we see the need for the following requirements:
\begin{itemize}
\itemindent=0.3in
\item
Profiles which are extensible to particular domains
\item
Protocols for remote access to data collections
\item
Query syntax and semantics for searching data collections
\item
File formats for data sets or subsets (e.g., XML document types)
\item
Flexible record formats for metadata describing data items
(e.g., XML document types)
\end{itemize}
In order to search across such a diversity of sources with a single query,
there must be some sort of ``common denominator", a common set of
search terms with shared semantics. 
The Dublin Core and related W3C efforts provide
this kind of standard for many kinds of ``document-like objects".
\cite{DC,W3Cmeta} 

However,
scientific data require metadata considerably beyond the Dublin Core.
For example, in addition to the Dublin Core categories, the ADIL supports
searches by:
\begin{itemize}
\itemindent=0.3in
\item
Sky position (e.g., in galactic coordinates)
\item
Astronomical Object (e.g., M31)
\item
Type of object (e.g., galaxy)
\item
Wavelength
\end{itemize}

Still other types of metadata are needed for 
archives of planetary data, such as orbital
positions and descriptions of atmospheres and clouds. 
While each discipline--and maybe even each project and 
instrument--may
require some unique metadata, we believe there is enough 
common ground to establish
standard profiles for broad classes of scientific data, just as the Dublin
Core has done for documents.  (See our proposals for astronomy metadata
in \cite{MPFP98,PMF98}.)

There must also be standards for the format of the results of
queries for data; a structured record that describes the kinds of objects
returned as data, such as images, tables, and datasets in various formats
such as FITS. 
USMARC records have served this role for many years
for bibliographic material, but new, more flexible record formats are needed
to support scientific data. 
The W3C RDF and schema initiative \cite{RDFSchema}
provides a sound framework for expressing such records, but it is critical
for communities to work to establish the appropriate standards.

\subsection{AML: Astronomical Markup Language, a metadata standard for astronomy}

The Astronomical Markup Language (AML) addresses the needs for standardized
metadata for Astronomy data. AML is an XML language describing various
kinds of data useful in astronomy, and is aimed at being an exchange format
for astronomical data, and especially metadata, over the Internet.
AML is both a proposed profile standard and a prototype implementation.
\cite{Guill98,GuillMur98}

Results of a search can be formatted as an AML document, that is, as
an XML document containing a description of the resource using the AML
DTD. The AML document can be processed by a program or presented
by a browser. 
Guillaume has created a Java applet to
browse AML documents as easily as one would browse HTML documents, but
with some additional features specific for astronomical data. 
For example, the
AML applet displays astronomical coordinates, and 
displays measurements
with the relevant units and uncertainties. 
\cite{Guill98,GuillMur98}

The use of AML is an improvement for both the information providers
and the users (who are astronomers). For the information providers,
XML separates the data from the user interface, so that different data
can be used with different user interfaces without any difficulty.
A small institute could also focus on the information, and let other institutes
provide user interfaces. For the users, the use of the AML browser
provides a uniform and unified way to access various data coming from different
servers. Finally, users wanting to get and process the data automatically,
can use the AML documents directly, as the AML browser applet does.
XML is much more useful for this purpose than HTML, because HTML
documents contain a mix of information about both the user interface and
the data.

The AML language is organized as seven types of objects. An AML
document is a collection of AML objects, describing different types of
information. AML objects may contain links to other AML objects,
and to external objects such as data, images, or documents.
The AML objects are summarized in Table 1. The AML language can be
easily extended, for example by adding a ``Set of images" object.

\begin{table*}
\caption{The objects defined by the Astronomical Markup Language}
\begin{tabular}{|l|p{4.0in}|} \hline
{\bf Object} & {\bf Definition} \\  \hline
Metadata & {An AML document is usually composed of the metadata part, and of one
the other parts.} \\ \hline

Astronomical object & 
{ This describes information about an astronomical object, with the identifiers
(the names), the coordinates, the object type, other information, a list
of measurements } \\ \hline
Article & {
This part only describes information about an article, including links
to the article, if available.
} \\ \hline
Table & {
Metadata for a table, and a link to the content of the table.
} \\ \hline
Set of tables (catalogue) &
{ A set of tables is a list of tables linked together, with information
about the set.
} \\ \hline
Image &
{Metadata about an image and the way it is stored, and a link
to the raw data of the image.
} \\ \hline
Person & 
{ Information about a person, usually an author of astronomical articles.
} \\ \hline
\end{tabular}
\end{table*}

AML records are designed to allow programs to automatically process and
analyze the metadata. Guillaume has demonstrated techniques
for automatically clustering astronomical information sources, e.g., applying
a graph partitioning algorithm to the keywords and links in AML records.
\cite{AMLMaps}
One outstanding feature of this work was that information from diverse sources
was successfully correlated, because the AML records are standardized.
It is easy to imagine how this work could be extended to support filtering
for specific users and selective dissemination of information.

\section{INTERACTION WITH THE DATA}

The search system described above locates information sources, and returns
AML records to describe them. From the AML records, the user identifies
data that appears to be of interest. There may in fact be a large
number of large datasets, so it is important for the user to select subsets
and subsamples from the data. For instance, it may be the case
that only one region or time period is required, or only certain measurements
are relevant. 

Sometimes the metadata itself is not sufficient
to make the selection, in which case the user needs to 
browse data itself. 
A low resolution ``thumbnail" image may be viewed, and the user may ask
to pan and zoom around the image, to examine in detail areas of possible
interest. Regions of an image may be selected with a bounding box,
data from tables may be examined, from which particular columns (fields)
and rows selected. It may be useful to make simple histograms or
other plots, to identify characteristics of the data, and it may be useful
to manipulate the color tables or other aspects of the display to highlight
features of the imagery. The dataset may contain tables of data,
or other multidimensional data structures, all of which need to 
be efficiently navigated in a similar fashion.

When the precise data of interest is identified, the user then requests
subsets and subsamples to be downloaded for detailed analysis. At
this point, the data will be input to data analysis programs or simulations.
These may range from simple graphing and spreadsheets, 
up to complex, multi-supercomputer
environments. In any case, the results may ultimately be published,
adding new documents and datasets to the library.

In the case of the ADIL, the FITS data might be filtered, combined with
data and models, and visualized. This might be done using AIPS++
or a similar package.\cite{AIPS++} The results of the analysis would be saved
as one or more FITS images, which might be entered in the ADIL when the
study is published in a journal. 
\cite{Plante97,PCM98}

Because of the elaborateness of scientific data, even the retrieval
step itself can sometimes involve fairly complex calculations above and
beyond the boolean matching typical of bibliographic data; e.g., applying
a pattern recognition algorithm to a database of images. Furthermore,
data may need to be processed and formatted even before it's browsed.
And finally, the scientific investigation may involve analysis of multiple
data sets to produce a composite data product distilled from diverse data
from several sources. Today, these types of activities are 
carried out routinely using a heterogeneous, ad-hoc assortment of applications,
typically specialized applications requiring access
to data on local disks. 
In the future, this will increasingly be done using ``workbenches"
(i.e., specialized Web portals such as \cite{biowb})
and ubiquitous computational GRIDs.
\cite{globus,FK98}

\section{A PROTOTYPE IMPLEMENTATION}

Over the past few years NCSA Project 30 has been constructing a
prototype which provides a sophisticated ``conversation with the data''
for astronomy data.
Our prototype uses NCSA Emerge and AML as the basis to build a system to
locate, browse, and retrieve astronomy data from the NCSA Astronomy Digital
Image Library and other data services. 

The data sources are already available
through standard Web interfaces which return HTML. We have added
the ability to use Z39.50 to query, installing the Gazelle Z39.50 gateway
on the data server if needed.

The Gazebo GUI implements a query construction interface, which presents
one or more profiles, i.e., standard sets of query terms and meanings.
The client configuration is loaded from the Gazebo gateway, so the same
client can have many ``views'' of the information space. The current
prototype implements both a ``simple'' query interface (a single list of
keywords), and an ``advanced'' interface (a graphical interface to construct
a boolean expression). The prototype supports a general purpose profile
for bibliographic searching, and a specialized astronomy profile.

The results of the query are returned as AML records, as well as HTML.
Creating AML (XML) records is usually a straightforward extension of the
existing code that generates HTML.

	The Gazebo GUI sends queries, encoded in the XML-based Gazebo
abstract search language, to the Gazebo gateway to be executed on a
set of target data sources.  Gazebo translates each query into the
native query syntax of each target data source, and executes it
remotely using the native search protocol of each target data source.
This behavior is highly configurable.  Requesting result records is
handled similarly; Gazebo translates the GUI's requests for results
into the target data sources' protocols.

	The result records returned by typical data sources are more
or less structured data.  Gazebo can return them unmodified or it can
process them through external CGI scripts which can translate them
from arbitrary file formats into any MIME type.  This is useful, for
instance, for providing HTML views of records in non-text formats.
The records are passed from the gateway to the
GUI, which displays them appropriately.

The GUI displays the number of records returned by each server, and
retrieves and shows the short records as requested. When a full record
is requested, the GUI retrieves the record and launches an appropriate
applet to display it. If the record is HTML, it is displayed with
the ICE HTML viewer. \cite{ICE} 
When the record is AML, the AML browser is invoked
to display it.

The AML record may contain pointers to abstracts and/or datasets.
The user may follow these links to view the actual data. The abstract
will be viewed with the HTML viewer. When a FITS dataset is selected,
the Horizon Image Browser\cite{Horizon} 
will be launched to browse the data, and
download it if desired.

\section{CONCLUSION AND RELATION TO OTHER WORK}

We have constructed a complete environment for locating astronomy information,
for examining and browsing metadata, and for browsing and accessing both
the text and the data. Our system is unique in that we support both
text and data, using a general, standards based protocol. We
have defined new protocols for describing astronomy data, and created a
much more complex ``conversation'' than most systems can support. The
flexible configuration and interoperable standards we use make it comparatively
easy to add databases.

It is important to reiterate that the Emerge software is extremely flexible,
and is used for several application communities. The astronomy specific
features are replaceable modules, the system can be customized for different
user communities.

The Gazebo gateway superficially resembles many conventional Web gateways
and portals. However, we use Z39.50 to distribute the queries and
AML to return the results. These standards assure much greater interoperability
than Web CGI and HTML.

Z39.50 has been widely used by libraries for many years, and there are
many efforts to federate Z39.50 services, such as the CIC Virtual
Electronic Library. \cite{CICVEL}
There is also a well established effort to 
standardized metadata for bibliographic
resources, e.g., the Dublin Core. \cite{DC} 
Our work is important
because it shows that Z39.50 can be used with scientific data. Our
protocol development and the AML extend the principles of the Dublin Core
to a significant body of scientific data.

The AML uses standard XML, but is not directly related to the still 
evolving
W3C metadata efforts.
\cite{W3Cmeta,RDFSchema}  
As RDF standards become established,
AML can presumably be aligned with them. For instance, the XML RDF schema
\cite{RDFSchema} and the XML-Data proposal \cite{XML-Data}
are likely to be important,
and the AML will follow these standards as they become established.

The Gazebo gateway and GUI implement a Query protocol using XML.
The W3C is currently in the early stages of defining a standard for representing
queries in XML.\cite{Valentine} 
When this standard matures, Gazebo will
support it appropriately.

\section{ ACKNOWLEDGEMENTS}
This work was partly funded by NASA Office of Space Science \cite{P30}, 
the National Science Foundation, 
and 
the National Center for Supercomputing Applications (NCSA)
at the University of Illi\-nois, Ur\-bana-Champaign.
NCSA is funded in part by
the National Science Foundation, the Advance Research
Projects Agency, corporate partners and the state and University
of Illinois.
Earlier parts of this work were funded by Project Horizon, a NASA
cooperative agreement. 
Some parts of this work were funded by the NSF/DARPA/NSF
Digital Library Initiative, and by the National Cancer Institute.

\end{document}